\documentstyle[12pt,aaspp4]{article}

\newcommand{\wtff}{W3/W4/W5/HB\,3}
\newcommand{\HI}{{\sc H\,i}}
\newcommand{\HII}{{\sc H\,ii}}
\newcommand{\etal}{et~al.\/}
\newcommand{\about}{$\sim$}
\newcommand{\dgr}{^\circ}
\newcommand{\fdgr}{^\circ\!\!.}
\newcommand{\lesim}{\,\raisebox{-0.5ex}{$\stackrel{\mbox{\footnotesize$<$}}
        {\mbox{\scriptsize$\sim$}}$}\,}

\begin{document}


\title {Radio polarimetric imaging of the interstellar medium: \\
magnetic field and diffuse ionized gas structure near the \wtff\ complex.}

\author{A.\,D.~Gray, T.\,L.~Landecker, P.\,E.~Dewdney}
\affil{National Research Council Canada, Herzberg Institute of Astrophysics,
Dominion Radio Astrophysical Observatory, Box~248, Penticton, BC, V2A~6K3,
Canada}

\author{A.\,R.~Taylor}
\affil{Department of Physics and Astronomy, University of Calgary,
2500~University Drive NW, Calgary, AB, T2N~1N4, Canada}

\author{A.\,G.~Willis}
\affil{National Research Council Canada, Herzberg Institute of Astrophysics,
Dominion Radio Astrophysical Observatory, Box~248, Penticton, BC, V2A~6K3,
Canada}

\author{M.~Normandeau\altaffilmark{1}}
\affil{Department of Physics and Astronomy, University of Calgary,
2500~University Drive NW, Calgary, AB, T2N~1N4, Canada}

\altaffiltext{1}{present address:Astronomy Department, University of
California, Berkeley, CA, 94720, USA}


\begin{abstract}
We have used polarimetric imaging to study the magneto-ionic medium of the
Galaxy, obtaining 1420\,MHz images with an angular resolution of $1'$ over
more than 40~square-degrees of sky around the \wtff\ \HII\ region/SNR
complex in the Perseus Arm.  Features detected in polarization angle are
imposed on the linearly polarized Galactic synchrotron background emission
by Faraday rotation arising in foreground ionized gas having an emission
measure as low as 1\,cm$^{-6}$\,pc.  Several new remarkable phenomena have
been identified, including: mottled polarization arising from random
fluctuations in a magneto-ionic screen that we identify with a medium in
the Perseus Arm, probably in the vicinity of the \HII\ regions themselves;
depolarization arising from very high rotation measures (several times
$10^3$\,rad\,m$^{-2}$) and rotation measure gradients due to the dense,
turbulent environs of the \HII\ regions; highly ordered features spanning
up to several degrees; and an extended influence of the \HII\ regions
beyond the boundaries defined by earlier observations.  In particular, the
effects of an extended, low-density ionized halo around the \HII\ region W4
are evident, probably an example of the extended \HII\ envelopes postulated
as the origin of weak recombination-line emission detected from the
Galactic ridge.  Our polarization observations can be understood if the
uniform magnetic field component in this envelope scales with the
square-root of electron density and is 20\,$\mu$G at the edge of the
depolarized region around W4, although this is probably an over-estimate
since the random field component will have a significant effect.
\end{abstract}



\section{Introduction}

Linearly polarized electromagnetic waves that pass through a magneto-ionic
medium (MIM; that is, a medium containing magnetic fields and free thermal 
electrons) undergoes a rotation of their polarization angle, known as Faraday
rotation.  Specifically, the amount $\Delta\phi$ (rad) by which the angle
is rotated at wavelength $\lambda$ (m) is $\Delta\phi=R_M\lambda^2$, where
$R_M$ is the rotation measure (rad\,m$^{-2}$), given by:
\begin{equation}
  R_M = K \int n_e \mbox{\boldmath $B$} \cdot d\mbox{\boldmath $l$}
  \label{eqrm}
\end{equation}
Here $K$ has a numerical value of 0.81 when the electron density $n_e$ is
in units of cm$^{-3}$, the magnetic field strength $B$ is in $\mu$G, and
the path length $l$ is in pc (see \cite{Lang74}).  A medium producing such
Faraday rotation is also referred to as a Faraday screen.

Diffuse ionized gas threaded by magnetic fields is ubiquitous in the
interstellar medium (ISM) of our Galaxy, so clues about the distribution
and properties of this MIM can be obtained by studying linearly polarized
signals from background emitters.  Measurements show that, in the general
ISM, $|R_M| \lesim\ 10^2$\,rad\,m$^{-2}$ (\cite{Clegg92}; values an
order of magnitude higher are not unknown, but are the exception rather
than the rule).  A range of detectable $\Delta\phi$ is then most readily
obtained at centi- and deci-metric wavelengths; at shorter wavelengths too
little rotation results, and at longer wavelengths too much rotation is
produced, yielding depolarization from the averaging of non-parallel
polarization vectors.

To date, radio observations of Galactic Faraday rotation effects in
linearly polarized emission have largely been limited to low spatial
resolution ($>0\fdgr5$), single-antenna studies of the diffuse Galactic
synchrotron component over wide areas of the sky
(see \cite{BS76}; \cite{Spoelstra84}), or to high spatial resolution
but targeted 
observations of polarized objects (mainly pulsars or extragalactic objects;
\cite{TI80}; \cite{SNKB81}; \cite{BMV88}; \cite{LS89};
\cite{Clegg92}).  In the presence of spatially 
varying rotation measure, either approach leads to limited information
about the detailed properties of the MIM component of the ISM, in the
former case because of the inherent low resolution and averaging effects
within a single beam, and in the latter case because of sparse sky
coverage.  The need for high resolution measurements covering a wide area
has been stressed by \cite{VS90}.

Any wide-scale study will, by necessity, have to rely on the polarized
fraction of the Galactic synchrotron component as the probe radiation.  The
non-thermal spectrum of this radiation favours longer wavelengths, where
the Faraday rotation produced by a given MIM is also greater.  Making
observations with resolutions of an arcminute or less at wavelengths longer
than a few centimetres requires the use of aperture synthesis techniques
(the Effelsberg 100\,m single-antenna telescope, for example, only achieves
sub-arcminute resolution at wavelengths shorter than $\lambda=3$\,cm; i.e.\
frequencies higher than 10\,GHz).

Recently, some interferometric observations have been made with the
Westerbork Synthesis Radio Telescope (WSRT) at $\lambda=92$\,cm (325\,MHz),
showing widely distributed polarized structures on scales from a few
arcminutes to several degrees across a number of $2\dgr$ fields at high
Galactic latitudes ($b>18\dgr$; \cite{Wieringa93}).  These structures
have no counterparts in total intensity, and are interpreted as arising
from sight-line dependent Faraday rotation effects in a local, foreground
MIM of low density.  In order to see ``through'' this local Faraday screen
to probe regions that are more distant and/or of higher column density it
is necessary to use a shorter wavelength.

We have used the Dominion Radio Astrophysical Observatory (DRAO) Synthesis
Telescope to produce interferometric images of polarization with arcminute
resolution at $\lambda=21$\,cm (1420\,MHz).  At this wavelength a rotation
measure 20 times as great is required to produce rotation angles comparable
to those seen at 92\,cm.  The region we have studied is the \wtff\ complex
(Figure~\ref{stokesI}).  W3, W4, and W5 are large, bright \HII\ regions at a
distance of 2.2\,kpc in the Perseus spiral arm of the Milky Way, related to
the OB association Cas~OB6.  They do not themselves emit polarized
radiation, but are rich in thermal electrons necessary for Faraday
rotation.  HB\,3 (G132.6+1.5; see \cite{Fesen95}) is a supernova
remnant (SNR)---and hence may itself be polarized---believed to be
interacting with the material in the vicinity of W3
(\cite{Caswell67}; \cite{Rout91}).  This region provides an
interesting environment 
in which Faraday rotation effects may be studied as polarized emission from
background sources, the diffuse Galactic emission, and HB\,3 pass through
the \HII\ regions and their environs.

Examination of our polarization data has revealed a range of remarkable
phenomena evident primarily in polarization angle, consistent with an
origin in the Faraday rotation effect.  We interpret these phenomena as
arising in a Faraday screen located in the Perseus arm of the Galaxy, and
see evidence for an extended influence of the \HII\ regions on the
surrounding ISM.  A presentation of our results and discussion of them in
relation to ISM studies is the topic of this paper.


\section{Data Acquisition and Reduction}

\subsection{The DRAO Synthesis Telescope}

The DRAO Synthesis Telescope (see \cite{Roger73} for a description of
the instrument prior to some recent upgrades) is a 7-element, east-west
aperture synthesis array located near Penticton, in British Columbia,
Canada.  The standard data product of the ST is described in some detail by
Normandeau, Taylor \& Dewdney (1997; hereafter referred to as NTD97), but,
briefly, it is comprised of simultaneous measurements of 408\,MHz
($\lambda=74$\,cm, 3\,MHz bandwidth) total radio continuum intensity
(see \cite{Veidt85}), 256-channel\footnote{128-channel at the time of the data
described in this paper.} 1420\,MHz ($\lambda=21$\,cm) neutral hydrogen
(\HI) spectroscopy (up to 4\,MHz bandwidth), and, of interest here,
1420\,MHz radio continuum (30\,MHz bandwidth) with full polarimetry.

A complete synthesis comprises twelve 12-hour observations in various
array configurations to provide to provide continuous baseline
coverage from 13 to 604\,m.  At 1420\,MHz the resulting images contain
information on spatial structures from approximately $1\dgr$ to $1'$ across
a field of about $2\dgr$ in diameter, with a noise-limited rms sensitivity
of 0.23\,mJy/beam.  In practice, the images are dynamic-range limited in the
vicinity of very strong sources.


\subsubsection{The 1420\,MHz Polarimeter System}

The ST antennas are fitted with orthogonal right- ($R$) and left- ($L$)
hand circularly polarized feed systems for 1420\,MHz operation.  The
continuum correlator provides all four cross-correlation products ($RR$,
$LL$, $RL$, $LR$), allowing recovery, after appropriate calibration
(see \cite{Smegal97}), of the four Stokes parameters, $I$, $Q$, $U$, $V$,
representing total intensity, two orthogonal components of linear
polarization, and circular polarization, respectively.  These four
parameters fully describe the polarization state of the incident radiation,
but in practice the $V$ data produced by the current system are dominated
by artefacts introduced by small ellipticity errors of the nominally
circular feeds, so it is mainly linearly polarized emission that is
studied.  This is not a major limitation since synchrotron sources---which
are the ones primarily studied with this instrument---do not emit
significant circular polarization, and the error introduced in linear
polarization is small.

After a complete synthesis observation the polarimeter precision is limited
to about 5\% in amplitude by calibration uncertainties, with polarization
angles determined to within $5\dgr$, although there is an additional
$3\dgr$ day-night variation of ionospheric origin that is not routinely
accounted for at present.  There is also an additive instrumental term of
approximately 0.3\% of $I$ on axis, rising quadratically to \about3\% of
$I$ at a radius of $1\dgr$ from the centre of the image.  Techniques are
presently being developed to reduce the magnitude of these various errors
and uncertainties, but all affect the data in this paper.  They do not,
however, affect the conclusions drawn from those data.


\subsection{The Observations}

The observations of the \wtff\ region presented in this paper were made in
June, July, November and December 1993 as a pilot study for the Canadian
Galactic Plane Survey.  Ten fields were observed, spanning an
$8\dgr\times6\dgr$ region centred near $l=134\fdgr7$, $b=+1\fdgr2$ (see
\ref{stokesI}).  Processing of the Stokes~$I$ and spectrometer data are
discussed in NTD97.  Additional processing applied to the polarimeter data
included polarization calibration to account for differing antenna
polarization responses, as well as calibration of polarization angle using
3C\,286, assumed to have a polarization angle of $33\fdgr5$ at 1420\,MHz.

The raw Stokes~$Q$ and $U$ images were affected by sidelobes arising
from the spurious instrumental response to W3, and also from the bright,
compact, off-field SNR 3C\,58 (G130.7+3.1, SN\,1181; see
\cite{Green88}, and references therein).  The individual fields were
therefore processed using {\tt MODCAL}, a visibility-based scheme for
removing such artefacts (see \cite{WH96}), applied in combination with a
standard {\tt CLEAN} algorithm for deconvolution of the synthesized beam.
The processed fields were then added together---with appropriate primary
beam weighting---to form composite images of the full pilot survey region.
Figures~\ref{stokesQ} and \ref{stokesU} show the composites of Stokes~$Q$
and $U$, respectively, while Figures~\ref{polint} and \ref{polang} show
polarized intensity ($P=\sqrt{Q^2+U^2}$) and polarization angle ($\theta_P
= \frac{1}{2} \arctan\frac{U}{Q}$).  Since the images of Stokes $Q$ and $U$
(and hence $P$ and $\theta_P$) do not include data on spacings shorter than
12.9\,m we will refer to them as ``interferometer images''.
Figure~\ref{overlay} shows the Stokes~$Q$ composite with overlaid contours of
2695\,MHz Stokes~$I$, to show the positions of the various sources in
relation to the detected polarization.


\section{Polarimetry as a Probe of the Interstellar Medium}
\label{intf}

To make use of polarimetry to probe the ionized component of the
interstellar medium requires an understanding of the interaction of
instrumental effects with the vector nature of polarization, particularly
in the presence of spatially varying and frequency-dependent Faraday
rotation.

Faraday rotation does not, in itself, alter the amplitude of the polarized
signal or its corresponding unpolarized portion.  The {\it observed}
polarized amplitude can, however, be altered by three depolarization
mechanisms.
\begin{enumerate}

\item {\it Bandwidth depolarization} occurs when the rotation measure is
sufficiently high that the Faraday rotation changes significantly across
the observing bandwidth, which intrinsically averages the resulting
non-parallel vectors.  For the DRAO Synthesis Telescope at 1420\,MHz, 50\%
depolarization occurs for $R_M=790$\,rad\,m$^{-2}$, and 99\% depolarization
occurs for $R_M=1250$\,rad\,m$^{-3}$ (assuming a perfectly rectangular
band-shape).

\item {\it Beam depolarization} occurs when large rotation-measure
gradients occur within the instrumental beamwidth, again resulting in the
averaging of disparate polarization vectors.

\item {\it Front-back depolarization} occurs when polarized emission
arises within a MIM, in which case polarized emission originating in
different regions will have different polarization angles, and vector
averaging will again take place (see \cite{Burn66}).

\end{enumerate}
For simplicity the last mechanism is not discussed further in this paper;
its inclusion would not change the conclusions, as the phenomena discussed
in this paper arise from propagation effects on a diffuse background, not
from intrinsic polarization within the MIM.

The above depolarization mechanisms mean that a MIM which produces observable
effects at a given wavelength must have a specific range of properties,
having a rotation measure that is neither too low to produce measurable
angle changes nor so high that it causes bandwidth depolarization.
Similarly, any spatial rotation-measure gradients cannot be too large on
scales small enough to cause beamwidth depolarization.  Faraday rotation,
even though it only directly affects polarization angle, can thus produce
structures in images of both observed polarized intensity {\it and}
polarization angle simply through depolarization.

There is a further effect to be considered with interferometer data.
Interferometers are inherently spatial filters, being unable to
image emission components that are distributed smoothly on large scales.
The exact cut-off depends on the instrument: with the DRAO Synthesis
Telescope we are able to detect structures up to about $1\dgr$ in size,
with a rapid decline in sensitivity to larger structures.  The effect of
this cut-off is readily understood for images of scalar quantities (e.g.\
Stokes~$I$), when it simply means that emission components on larger scales
are absent from the images.  The situation is slightly more complex,
however, for a vector quantity like polarization.

Consider the case of a large-scale, uniform polarization field, i.e.\
constant polarized intensity and polarization angle, and hence constant
Stokes~$Q$ and $U$.  Attempting to image such a field with an
interferometer would result in a non-detection, since all of the emission
is on large scales.  If this polarization field first passes through a
non-uniform MIM that produces spatially varying Faraday rotation then the
polarization angle, and hence $Q$ and $U$, will vary according to the local
properties of the MIM.  The polarized intensity, which is the quadrature
sum of $Q$ and $U$, will still be uniform, since Faraday rotation does not
affect intensity.  However, spatially filtering this distribution by
observing it with an interferometer will result in detected signal only
where the variations in $Q$ and $U$ occur on scales to which the
interferometer is sensitive.  That is, apparent structure will be induced
in polarized intensity due solely to variations in polarization angle.
Since Faraday rotation depends on both electron density and magnetic field
orientation, both of which vary on small scales, there may be smaller-scale
variations visible in the vector polarization than in the scalar total
intensity (Stokes $I$) data.  The interferometer is thus a valuable tool
for studies of ISM polarization properties on small-scales.

It is important to note that, since the vector field effectively subtracted
by spatial filtering is not spatially uniform, the interferometer image
will not preserve the difference in polarization angle measured between two
widely separated points in the image.  However, provided that any such
difference is measured over scales small compared to the largest structure
that the interferometer can image, the error introduced will be negligible,
since the subtracted vector field does not vary rapidly on such scales.


\section{Results}

Several notable features are seen in the polarization images, including:
  \begin{enumerate}

    \item a mottled pattern, most prominent south of \wtff,
          but present over much of the field shown here;

    \item an absence of detected polarization on lines-of-sight towards
          the \HII\ regions; and

    \item an elliptical feature roughly coincident with W5.

  \end{enumerate}
All of these features are seen in raw images from the telescope, and are
consistent in regions where the fields overlap.  All are also seen in---and
are consistent with---new data for this region obtained for the Canadian
Galactic Plane Survey project, which used different pointing centres from
the data presented here.

In this paper we concern ourselves with the details of features~1 and 2,
which we will show are likely to be manifestations of the ISM in the
Perseus arm, in which \wtff\ reside.  Feature~3 points to a more local
effect in the ISM, since it is superimposed on the depolarized zone
(Feature~2) of W5.  This phenomenon is mentioned in the following
discussion, but is examined in detail elsewhere (\cite{Gray98}).


\subsection{Mottled Polarization Structures}
\label{mottle}

The images of Stokes~$Q$ and $U$ (Figures~\ref{stokesQ} and \ref{stokesU})
display a widespread mottled pattern.  The pattern extends to the edge of
the observed region, but is most prominent to the south of W3/W4/W5.
Defining the ``cell size'' as the scale-size over which polarization angle
variations of $180\dgr$ occur, the cells range from \about$1'$ close to
W4 to \about$10'$ farther away.  Elsewhere they range up to several tens
of arcminutes in size.  Similar structures are seen in essentially all
images made of other regions in and near the Galactic plane, so this is not
just a peculiarity of the \wtff\ region.

The mottling is reflected strongly in polarization angle $\theta_P$
(Figure~\ref{polang}), but there is only weak structure in polarized intensity
$P$ (Figure~\ref{polint}), and no corresponding Stokes $I$ emission.  If the
mottled $P$ structure represents some component with varying $P$ then its
absence in $I$ would require that the polarized fraction $P/I$ also vary to
keep $I$ constant.  Since, with few exceptions, there is also no structure
in $P$ that is not in $\theta_P$, and the $P$ structure is most prominent
where $\theta_P$ varies most rapidly, the simplest explanation is that, as
described in \S\ref{intf}, the $P$ structure here is merely that expected
in interferometer data from a pure polarization angle effect.  That is, the
underlying phenomenon is indeed Faraday rotation.  The widespread nature of
the polarization seen here indicates that the radiation must originate in
the diffuse Galactic background emission, and is being acted on by some
foreground screen of MIM.

It is possible to place some constraints on the location of the Faraday
screen.  An upper limit to the distance to the screen is obtained by
considering the total 1420\,MHz diffuse background emission in this region,
which has a brightness temperature of 4.2\,K (see \cite{KR80}).  This will
include some thermal emission at this frequency, but assuming a polarized
fraction of 35\% (\cite{Spoelstra84}) yields an upper limit on the polarized
emission of 1.5\,K.  Roger (1969) measured synchrotron emissivity
between W4 and the Sun at 22\,MHz and found that 29\% of the total emission
originates behind W4, corresponding to at most 0.45\,K of polarized
emission at 1420\,MHz.  We detect up to 0.3\,K of diffuse polarized
intensity in our images, which is a lower limit on the total signal passing
through the screen, since large-scale structure is not detected by the
interferometer.  Since the synchrotron emissivity of the Galaxy is
concentrated in the spiral arms (\cite{BKB85}), this suggests that the screen
cannot be located much farther away than W3/W4/W5, or there would be
insufficient polarized signal behind it to account for our observations.

A lower limit for the distance derives from the fact that the 71\% (or up
to 1.1\,K) of the polarized emission that arises in front of W4 is not
acted on by the screen, otherwise we would see it in the depolarized zones
coincident with the \HII\ regions (discussed in \S\ref{zod} below).  The
screen must therefore lie behind this emission.  But where does this
emission originate?  \cite{WS74} argue that the strong, ordered
polarization detected towards the \wtff\ region in single-antenna
measurements arises within 500\,pc of the Sun.  The strength of this
polarized signal at 1411\,MHz is about 0.5\,K (see \cite{BS76}), a lower limit
since single-antenna measurements will tend to underestimate the signal
strength due to beam depolarization.  The location of the remainder---up to
0.6\,K---is likely to be closer to the Perseus arm since the synchrotron
emissivity is concentrated in the arms (\cite{BKB85}).  We know that the
screen does not act on this component, and also that the thermal electrons
necessary to produce the Faraday rotation are also concentrated in the
spiral arms (\cite{TC93}).  It is therefore most likely that the screen is
located in the Perseus arm itself.

This screen may be an extended halo around the \HII\ regions.  In \S4.2 we
present a model of the electron density profile of the halo around W4.  If
this halo is parameterized as being $100L$\,pc in size ($L$ is of order 1),
then the thermal electron density in the tail of the Gaussian model is
$n_e=0.3/L$\,cm$^{-3}$.  The magnetic field in the Perseus arm has a
line-of-sight component of $B_{||} = 2.5\,\mu$G, with variations $\delta
B/B=1.4$ (\cite{ARS88}).  For this electron density and magnetic field
variation a change in polarization angle between adjacent sight-lines of
$180\dgr$ requires a path of length $80L$\,pc.  That is, the depth of the
screen needs to be essentially the same as the depth of the W4 halo.


\subsection{Depolarized Zones Around \HII\ Regions}
\label{zod}

While the mottled polarization is detected over large areas, closer to the
\HII\ regions a different behaviour is seen in our data.  Across the face
of W3 and W4 and on sight-lines immediately adjacent to them there is no
apparent polarization (Figures~\ref{stokesQ} through \ref{polang}; a small
amount of instrumental polarization is visible in these images as faint
``ghosts'' of the total intensity emission).  We attribute the lack of
observed polarization to high rotation measures and rotation measure
gradients on small-scales in the dense, turbulent environment of the \HII\
regions, resulting in a strong spatial and frequency dependence of
polarization angle.  Beam and bandwidth depolarization effects then
dominate, with the result that the detected signal falls to zero.  Some
weak polarization is seen from the non-thermal emission from the SNR,
HB\,3, but large portions appear unpolarized, possibly as a result of
depolarization by the adjacent \HII\ regions.

As noted by Braunsfurth (1983), both W3 and W4 are embedded in an extended halo
of diffuse radio continuum emission; this is visible in single-antenna,
Stokes~$I$ maps of this region at 1420\,MHz (\cite{KR80}) and 2695\,MHz
(\cite{Furst90}; see Figure~\ref{stokesI}).  The depolarized zone around W4
is particularly clear to the south, where the zone boundary closely follows
a contour of total intensity (see Figure~\ref{depol}).  Since the halo is not
visible in the Stokes~$I$ interferometer images, it must not have
significant structure on scales smaller than about $1\dgr$.  Examination of
the radial profile of the halo at 2695\,MHz to a distance of about $3\dgr$
south-west of the centre of the W4 loop, a region relatively free from
discrete emission, shows that its rise is well approximated by a Gaussian
(temperature $T$/mK$=550\,e^{-(r/78')^2} + 30$, for $r$ in arcminutes)
until within about $1\dgr$ of the centre of W4, after which it rises much
more sharply (Figure~\ref{rise}).  It is at this point, which we refer to as
the ``vanishing point'', that the polarization also disappears sharply,
strongly suggesting that this is a related phenomenon.

Modelling the halo around W4 as a spherical Gaussian yields an electron
density of $n_e=2.8$\,cm$^{-3}$ at the vanishing point, corresponding to
thermal electron column density of $3.5\times10^2$\,cm$^{-3}$\,pc.  The
interstellar magnetic field strength varies as a power-law with electron
density, with index $\frac{1}{3} < \alpha < \frac{2}{3}$ (see \cite{Go88}).
The Faraday rotation produced by such a cloud would vary in a Gaussian
fashion with radius, and hence the observed polarization angle will also
vary with radius, ``wrapping'' at $\pm180\dgr$ with some spatial interval
that will depend on the magnitude of the Faraday rotation.  In the case of
our data, wrapping on scales smaller than $1'$ would produce beam
depolarization, so we require that the scale-size of the wrapping be no
smaller than $1'$ at the vanishing point, otherwise the polarization would
be seen to disappear outside the brightest emission from W4 rather than at
its boundary.  This condition is met for a line-of-sight magnetic field at
the vanishing point of approximately $20\,\mu$G.  This value does not
depend strongly on $\alpha$ within the range quoted above, but random
fluctuations (\S\ref{mottle}) confuse the radial variation of polarization
angle in our data, so this value is an upper limit.  However, we note that
the bandwidth depolarization effects produced by the resulting Faraday
rotation are also consistent with our observations, and the spatial
interval between polarization angle wrap points would be about $25'$ at a
radius of $3\dgr$, which may mean that the bar-like structure at the bottom
of Figure~\ref{polang} is related to the halo.

So far we have considered only the W3/W4 region; the case of W5 is
complicated by the presence of the elliptical polarization feature.  This
is discussed in detail by Gray et al.\ (1998), who conclude that it is a
foreground phenomenon, probably arising from a MIM in the inter-arm region.
Nonetheless, it is superimposed on a region of low or absent polarization,
consistent with a depolarized zone coincident with W5, but there is no
obvious ``vanishing point'' associated with W5, which lacks the extended
halo of W4 (\cite{Br83}; see also Figure~\ref{stokesI}).


\section{Distribution of the Faraday Screen}

Although the limited region presented here does not offer any direct
constraint on the overall extent of the Faraday screen in either latitude
or longitude, there are some qualitative indications in other observations.


\subsection{Comparison With Other Data}

DRAO 1420\,MHz data that we obtained for several of the
Wieringa et al.\ high latitude ($b=18$--$52\dgr$) fields showed little
significant polarization, and none similar to the structures observed by
Wieringa et al.\ at 325\,MHz.  If these structures were intrinsic to
the polarized emission we should still be able to detect them at 1420\,MHz
despite the falling spectrum of Galactic background emission, particularly
with the decreased Faraday rotation and associated depolarization effects.
However, if we assume that these structures are {\it due} to Faraday
rotation then our result is entirely expected, and indeed supports the
Wieringa et al.\ model of this phenomenon, namely that the structures
seen at 325\,MHz are due to a local, foreground screen.

Other high latitude fields observed at DRAO at 1420\,MHz also show few
significant polarization structures, but polarization structures apparently
unrelated to Stokes~$I$ emission appear essentially ubiquitous in fields
observed for the Canadian Galactic Plane Survey project as well as other
fields observed near the Galactic plane in the course of normal DRAO
operations.  It is clear, then, that the structures of interest at
1420\,MHz are generally confined to lower latitudes ($b\lesim10\dgr$), and
hence the MIM responsible for their formation must be confined to the disc
component of the Galaxy.  The general properties of these structures are
similar to those detected in the WSRT 325\,MHz observations, but they
cannot be manifestations of the same MIM, since depolarization effects will
become important at 325MHz by the time the detectable variations are
present at 1420MHz.  For example, a change in rotation measure of $\Delta
R_M=2$\,rad\,m$^{-2}$ will produce a change in polarization angle of only
$\Delta\theta_P=5\dgr$ at 1420\,MHz, which, when averaged within a beam,
will have a negligible depolarizing effect, but at 325\,MHz
$\Delta\theta_P=97\dgr$, which will produce essentially complete
depolarization.  Similarly, at the narrowest band used at 325\,MHz
(2.3\,MHz) there would be total bandwidth depolarization for a rotation
measure of approximately $R_M=260$\,rad\,m$^{-2}$, which would produce only
6\% depolarization in our data.  The difference in latitude extent observed
at 325\,MHz and 1420\,MHz further suggests that the media being probed are
not only distinct but lie at different distances.


\subsection{Environment of \HII\ Regions}

In our data there is also some evidence of a longitude dependence in both
the structures themselves and in their latitude extent.  Indications in the
data at hand, both in the \wtff\ region and elsewhere along the Galactic
plane, are that the presence of large \HII\ regions may be a factor in
enhancing the polarization structures and increasing their latitude extent.
Since the detection of the polarized structures depends on thermal electron
density, this suggests that the \HII\ regions might be responsible for a
surrounding region of weakly enhanced levels of thermal electrons---that
is, enough to cause significant rotation measures, but of sufficiently low
emission measure that they are not detected in total intensity data---in
the ISM over an area several times larger than the nominal boundary of the
\HII\ region.  We may be detecting the effect of the extended \HII\
envelopes (EHEs) proposed by Anantharamaiah (1985, 1986) to explain widely
distributed 325\,MHz radio recombination-line (RRL) emission from the
Galactic ridge.  These RRLs have not previously been localized sufficiently
well that they could be associated with known \HII\ regions (see \cite{Ka89}),
but appear to originate in gas with densities and path lengths (\cite{An86})
comparable to those we propose for the Faraday screen.  Our Faraday
rotation technique may thus provide the link required to establish this
connection.  As more data become available from the Canadian Galactic Plane
Survey we will gain a clearer insight into this issue.


\section{Conclusion}

The DRAO Synthesis Telescope's 1420\,MHz polarimeter system has provided us
with a powerful new tool for probing the interstellar medium over wide
fields, yielding information on the structure of the MIM that is not
obtainable by other means.  The study presented here has shown widespread
Faraday rotation effects in the background Galactic synchrotron emission in
the direction of the \wtff\ complex, with several interesting phenomena
which can be associated with very low densities of thermal electrons
(emission measures as low as 1\,cm$^{-6}$\,pc) threaded by the interstellar
magnetic field.  These include:
\begin{enumerate}

\item a wide-spread mottled polarization structure that we attribute to
the action of a widely distributed screen, which is constrained to lie 
in the Perseus arm, probably in the immediate vicinity of the the \HII\
regions themselves.

\item depolarization zones around the \HII\ regions, arising due to very
high rotation measures of at least several times $10^3$\,rad\,m$^{-2}$ and
spatial rotation measure gradients.

\item the suggestion of an extended region of influence around the \HII\
regions, well beyond the apparent boundary of their emission evident in
radio and other wave-bands, potentially providing an observational
confirmation of the extended \HII\ envelopes postulated by Anantharamaiah
(1985, 1986).

\item the elliptical feature coincident with W5 which, although not
discussed in detail here (see \cite{Gray98}), is probably a foreground
object seen against the depolarized zone of the \HII\ region.  This leads
to the important observation that {\it regular polarization structures may
be widespread, but their existence will generally be masked by the presence
of random fluctuations in the ISM.}

\end{enumerate}
One important aspect of this work is that it demonstrates the possibility
of using objects of known distance to estimate the distribution of the
polarized emission as well as the MIM along the line-of-sight.  It is also
important to note that the $\lambda^2$ dependence of Faraday rotation means
that substantial differences in the MIM properties are necessary to produce
comparable results at widely differing wavelengths.  Data at other, nearby
wavelengths may also be useful in establishing the absolute rotation
measure, and hence the absolute properties of the MIM (the data at hand
constrain the variations in the MIM properties, but not the underlying
distribution).  Data at more disparate wavelengths will provide information
about regions of lower or higher density.  Multi-wavelength studies of this
type are thus important in obtaining information about the ionized
component of the ISM and the interstellar magnetic fields in our Galaxy.


\section*{Acknowledgements}

The Dominion Radio Astrophysical Observatory is operated as a national
facility by the National Research Council Canada.  The Canadian Galactic
Plane Survey is a Canadian project with international partners, and is
supported by a grant from the Natural Sciences and Engineering Research
Council of Canada.  We also thank an anonymous referee for comments helpful
in the preparation of the final manuscript.



\clearpage

\figcaption{Single-antenna image of the \wtff\ region, taken from the
2695\,MHz total intensity survey of F\"urst \etal\ (1990).  The sources
mentioned in the text are labelled.  Owing to the falling synchrotron
spectrum, images at this frequency highlight thermal emission.  The
greyscale has been deliberately saturated at 0.3\,K, with contours at 0.3,
0.6, 1.2, 2.4, 4.8, 9.6, and 19.2\,K to show the higher intensities.  Note
the extensive halo around W3 and W4, which is absent from W5 [the horn-like
projections in this halo extending upwards from W3 and the eastern edge of
W4 are part of the ``chimney'' phenomenon reported by Normandeau,
Taylor \& Dewdney (1996)].
\label{stokesI}}

\figcaption{Composite interferometer image of Stokes~$Q$ for the
pilot survey fields.  The greyscale runs from $-0.25$ to $+0.25$\,K, with
lighter shades of grey indicating higher temperatures.  Some residual
instrumental polarization is present in these images, giving strong
unpolarized sources (e.g.\ W3, at centre-right) the appearance of having
weak polarized emission.
\label{stokesQ}}

\figcaption{Composite interferometer image of Stokes~$U$ for the
pilot survey fields.  The comments for Figure~\ref{stokesQ} also apply
here.
\label{stokesU}}

\figcaption{Composite image of polarized intensity for the pilot
survey fields.  The comments for Figure~\ref{stokesQ} also apply here, except
that the greyscale runs from 0.0 to 0.35\,K.  The primary beam correction
applied to these data boosts the noise towards the edges of the individual
fields, giving rise to the ``honeycomb'' appearance seen here.
\label{polint}}

\figcaption{Composite images of polarization angle for the pilot
survey fields.  The comments for Figure~\ref{stokesQ} also apply here, except
that the greyscale runs from $-90\dgr$ to $+90\dgr$.
\label{polang}}

\figcaption{The same greyscale as Figure~\ref{stokesQ}, but with overlaid
contours of 2695\,MHz Stokes~$I$ F\"urst \etal\ (1990) at 0.3, 0.6,
1.2, 2.4, 4.8, 
9.6, and 19.2\,K, showing the relative locations of the Stokes~$I$
emission.
\label{overlay}}

\figcaption{As for Figure~\ref{overlay}, magnified to show the region
around W4 in greater detail.  Note how the limit of the detected
polarization follows a line of constant intensity---the lowest contour at
0.3\,K---which lies well outside the boundary of the strongest emission.
Even small-scale irregularities in this contour are apparently followed by
the limit of detected polarization.
\label{depol}}

\figcaption{Plot showing the sky brightness temperature as a function of
radius from the centre of W4.  The two dashed curves are data averaged over
two sectors, each $10\dgr$ wide, centred at position angle $210\dgr$ and
$220\dgr$ (i.e. extending south-west of W4).  The solid curve is a Gaussian
approximation to the rise at large radii (see text).  Within about $60'$
radius the sky temperature rises more steeply than the Gaussian.  It is
within this region that the polarization is no longer detected.
\label{rise}}

%
%
%
%
%
%
%
%
%
%
%
%
%
%
%


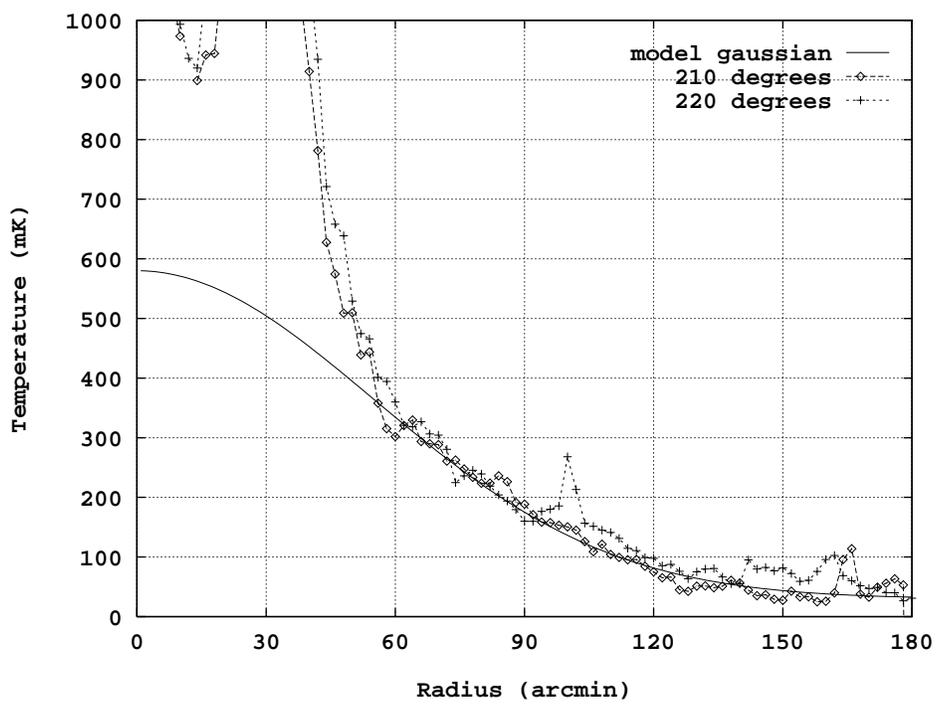
\end{document}